\documentclass[10pt]{article}
\usepackage{graphicx}
\usepackage{amsmath}
\usepackage{amssymb}
\usepackage{caption2}
\begin{document}
\bibliographystyle{prsty}
\begin{center}
{\large {\bf \sc{  Analysis of mass modifications of the vector and axialvector heavy mesons in the nuclear matter with the QCD sum rules  }}} \\[2mm]
Zhi-Gang Wang \footnote{E-mail,wangzgyiti@yahoo.com.cn.  }\\[2mm]
 Department of Physics, North China Electric Power University, Baoding 071003, P. R. China
\end{center}

\begin{abstract}
In this article, we calculate the   mass modifications  of the vector and axialvector mesons $D^*$, $B^*$, $D_1$ and
$B_1$ in the nuclear matter with the QCD sum rules, and obtain  the mass-shifts $\delta M_{D^*}=-71\,\rm{MeV}$, $\delta M_{B^*}=-380\,\rm{MeV}$,
$\delta M_{D_1}=72\,\rm{MeV}$,  $\delta M_{B_1}=264\,\rm{MeV}$, and the scattering lengths   $a_{D^*}=-1.07\,\rm{fm}$,
$a_{B^*}=-7.17\,\rm{fm}$, $a_{D_1}=1.15\,\rm{fm}$ and $a_{B_1}=5.03\,\rm{fm}$ for the $D^*N$, $B^*N$, $D_1N$ and $B_1N$ interactions, respectively.
\end{abstract}



\section{Introduction}

The  modifications of the hadron properties in the nuclear matter
   can affect  the productions  of the open-charmed mesons and  the $J/\psi$ in
the relativistic heavy ion collisions,   the higher charmonium states, such as the $\psi'$, $\chi_{c1}$, $\chi_{c2}$, etc,  are
considered as the major source of the $J/\psi$  \cite{Jpsi-Source}.
The charmed mesons can  obtain mass augments or reductions in the nuclear matter,
if the mass modifications are large enough, the decays of the higher  charmonium states to the
charmed meson pairs can be facilitated or suppressed remarkably due to the available  phase-space,
thus the decays to  the lowest  state $J/\psi$ are greatly modified \cite{H-to-Jspi}. For example,
 the higher  charmonium states can decay to the $D\bar{D}$ pairs instead
of decaying to the lowest  state $J/\psi$, if the  mass reductions
   of the $D$ and $\bar{D}$ mesons  are large enough. On the other hand, the suppression of the $J/\psi$ production in the relativistic  heavy
ion collisions is considered as an important
signature to identify the possible phase  transition  to the
quark-gluon plasma \cite{Matsui86}. We should be careful before making  definite conclusions.

   The QCD sum rules is a powerful theoretical tool  in
 studying the in-medium hadronic properties \cite{SVZ79}, and has been applied extensively
  to study the light-flavor hadrons and  charmonium states in the nuclear matter \cite{C-parameter,Drukarev1991,Jpsi-etac}. The
  works on the heavy mesons and heavy baryons are  few, only the $D$,  $B$, $D_0$, $B_0$,  $\Lambda_c$,  $\Lambda_b$, $\Sigma_c$ and
    $\Sigma_b$  are studied with the QCD sum rules \cite{Hayashigaki,WangHuang,DB-SR,SumRule-D0,SumRule-VA,WangH}.
The heavy mesons  contain a  heavy quark and a light quark,
the existence of a light quark in the heavy mesons leads to large
difference between the  mass-shifts of
 the heavy mesons and heavy quarkonia  in the nuclear matter.
The former have  large contributions  from the light-quark
condensates,  while the latter
  are dominated by the gluon condensates \cite{Jpsi-etac,Hayashigaki,WangHuang,DB-SR,SumRule-D0}.
  In this article, we study the mass modifications of the vector mesons $D^*$, $B^*$ and axialvector mesons $D_1$, $B_1$ in the nuclear
matter using the QCD sum rules. The present predictions can be confronted with the experimental data from the
CBM and PANDA collaborations  in the future  \cite{CBM,PANDA}.

The article is arranged as follows:  we study the mass modifications of the vector and axialvector  mesons $D^*$, $B^*$, $D_1$ and $B_1$ in the nuclear matter
 with  the  QCD sum rules in Sec.2; in Sec.3, we present the numerical results and discussions; and Sec.4 is reserved for our
conclusions.

\section{Mass modifications of the $D^*$, $B^*$, $D_1$ and $B_1$ in the nuclear matter with  QCD sum rules}

We study the mass modifications of the $D^*$ and $D_1$ mesons in nuclear matter
with the two-point correlation functions $\Pi_{\mu\nu}(q)$,
\begin{eqnarray}
\Pi_{\mu\nu}(q) &=& i\int d^{4}x\ e^{iq \cdot x} \langle T\left\{J_\mu(x)J_\nu^{\dag}(0)\right\} \rangle_{\rho_N} \, ,
 \end{eqnarray}
 where the $J_\mu(x)$ denotes the isospin averaged currents $\eta_\mu(x)$ and $\eta_{5\mu}(x)$,
\begin{eqnarray}
 \eta_\mu(x) &=&\eta_\mu^\dag(x) =\frac{\bar{c}(x)\gamma_\mu q(x)+\bar{q}(x)\gamma_\mu c(x)}{2}\, , \nonumber\\
  \eta_{5\mu}(x) &=&\eta_{5\mu}^\dag(x) =\frac{\bar{c}(x)\gamma_\mu \gamma_5q(x)+\bar{q}(x)\gamma_\mu\gamma_5 c(x)}{2}\,,
\end{eqnarray}
 which interpolate the vector and axialvector mesons $D^*$ and $D_1$, respectively,  the $q$ denotes the $u$ or $d$ quark.
 The $\bar{c}q$ and $\bar{q}c$ mesons maybe obtain
different mass modifications in the nuclear matter,
   just like the $K^+$ and $K^-$ mesons \cite{KaonK}, for example, Hilger et al observe  that there
   exist particle-antiparticle mass splittings  for the scalar and pseudoscalar mesons \cite{DB-SR,SumRule-D0}.
  In this article, we intend to study whether or not the decays of the higher
charmonium states to the   $D^*\bar{D}^*$ and $D_1\bar{D}_1$ states are facilitated in the phase-space, and prefer the average values as the
particle-antiparticle mass splittings cannot modify the total mass of the particle-antiparticle pair.

At the low nuclear density,  the in-medium condensates $\langle{\cal {O}}\rangle_{\rho_N}$,
\begin{eqnarray}
\langle{\cal{O}}\rangle_{\rho_N} &=&\langle{\cal{O}}\rangle+\frac{\rho_N}{2M_N}\langle {\cal{O}}\rangle_N \, ,
\end{eqnarray}
based on the Fermi gas model, where the   $\langle{\cal{O}}\rangle$ and $\langle
{\cal{O}}\rangle_N$ denote the vacuum condensates and nuclear matter induced condensates,  respectively, the  $\rho_N$ is the density of the nuclear matter \cite{Drukarev1991}.
Accordingly, the correlation functions  $\Pi_{\mu\nu}(q)$ can be divided
into  a vacuum part $\Pi^0_{\mu\nu}(q)$  and a static one-nucleon part
$T^N_{\mu\nu}(q)$  in the Fermi gas
approximation for the nuclear matter \cite{Drukarev1991,Hayashigaki},
\begin{eqnarray}
\Pi_{\mu\nu}(q) &=&\Pi^{0}_{\mu\nu}(q)+ \frac{\rho_N}{2M_N}T^{N}_{\mu\nu}(q)\, ,
 \end{eqnarray}
where
\begin{eqnarray}
T^{N}_{\mu\nu}(\omega,\mbox{\boldmath $q$}\,) &=&i\int d^{4}x e^{iq\cdot x}\langle N(p)|
T\left\{J_\mu(x)J_\mu^{\dag}(0)\right\} |N(p) \rangle\, ,
\end{eqnarray}
the $|N(p)\rangle$ denotes the isospin and
spin averaged static nucleon state with the four-momentum $p =
(M_N,0)$, and normalized as  $\langle N(\mbox{\boldmath $p$})|N(\mbox{\boldmath
$p$}')\rangle = (2\pi)^{3} 2p_{0}\delta^{3}(\mbox{\boldmath
$p$}-\mbox{\boldmath $p$}')$ \cite{Hayashigaki}. The $T^{N}_{\mu\nu}(q)$ happen  to be the
current-nucleon forward scattering amplitudes.
We can decompose the correlation functions $T^{N}_{\mu\nu}(\omega,\mbox{\boldmath $q$}\,)$ as
\begin{eqnarray}
T^{N}_{\mu\nu}(\omega,\mbox{\boldmath $q$}\,) &=&T_{N}(\omega,\mbox{\boldmath $q$}\,)\left(-g_{\mu\nu}+\frac{q_\mu q_\nu}{q^2}\right)+\Pi^0_N(\omega,\mbox{\boldmath $q$}\,) \frac{q_\mu q_\nu}{q^2} \, ,
\end{eqnarray}
according to Lorentz covariance, where the $T_{N}(\omega,\mbox{\boldmath $q$}\,)$ denotes the contributions from the vector and axialvector mesons,
 and the $T^0_{N}(\omega,\mbox{\boldmath $q$}\,)$ denotes the contributions from the scalar and pseudoscalar mesons. The interpolating
 currents $\eta_\mu(x)$ and $\eta_{5\mu}$  have non-vanishing couplings  with the scalar and pseudoscalar mesons $D_0$ and $D$, respectively, i.e. $\langle 0|\eta_\mu(0)|D_0+\bar{D}_0\rangle =f_{D_0}q_\mu$ and $\langle 0|\eta_{5\mu}(0)|D+\bar{D}\rangle =if_{D}q_{\mu}$. We can exclude the contaminations by choosing the tensor structure $-g_{\mu\nu}+\frac{q_\mu q_\nu}{q^2}$,   and only take into account the vector and axialvector mesons through the definitions,
\begin{eqnarray}
\langle 0|\eta_\mu(0)|D^*+\bar{D}^*\rangle &=&f_{D^*}M_{D^*}\epsilon_\mu\,,\nonumber\\
\langle 0|\eta_{5\mu}(0)|D_1+\bar{D}_1\rangle &=&f_{D_1}M_{D_1}\epsilon_{\mu}\,,
\end{eqnarray}
with summations of the polarization vectors  $\sum_\lambda \epsilon_\mu(\lambda,q)\epsilon^*_\nu(\lambda,q)=-g_{\mu\nu}+\frac{q_\mu q_\nu}{q^2}$.

In the limit of the $3$-vector $\mbox{\boldmath $q$}\rightarrow {\bf 0}$, the
correlation functions $T_{N}(\omega,\mbox{\boldmath $q$}\,)$ can be related to the $D^*N$ and $D_1N$
scattering $T$-matrixes, i.e. ${{\cal T}_{D^*N}}(M_{D^*},0) =
8\pi(M_N+M_{D^*})a_{D^*}$ and ${{\cal T}_{D_1N}}(M_{D_1},0) =
8\pi(M_N+M_{D_1})a_{D_1}$, where the  $a_{D^*}$ and $a_{D_1}$ are the $D^*N$ and $D_1N$
scattering lengths, respectively. Near the pole positions of the $D^*$ and $D_1$ mesons, the
phenomenological spectral densities $\rho(\omega,0)$ can be
parameterized with three unknown   parameters $a,b$ and $c$
\cite{Hayashigaki},
\begin{eqnarray}
\rho(\omega,0) &=& -\frac{f_{D^*/D_1}^2M_{D^*/D_1}^2}{\pi} \mbox{Im} \left[\frac{{{\cal T}_{D^*/D_1N}}(\omega,{\bf 0})}{(\omega^{2}-
M_{D^*/D_1}^2+i\varepsilon)^{2}} \right]+ \cdots \,, \\
&=& a\,\frac{d}{d\omega^2}\delta(\omega^{2}-M_{D^*/D_1}^2) +
b\,\delta(\omega^{2}-M_{D^*/D_1}^2) + c\,\delta(\omega^{2}-s_{0})\, ,
\end{eqnarray}
 the terms
denoted by $\cdots$ represent the continuum contributions. The first
term denotes the double-pole term, and corresponds  to the on-shell (i.e. $\omega^2=M_{D^*/D_1}^2$)
effects  of the $T$-matrixes,
\begin{eqnarray}
a&=&-8\pi(M_N+M_{D^*/D_1})a_{D^*/D_1}f_{D^*/D_1}^2M_{D^*/D_1}^2\, ,
\end{eqnarray}
  and related with the mass-shifts of the
 $D^*$ and $D_1$ mesons  through the relation
 \begin{eqnarray}
\delta M_{D^*/D_1} &=&-\frac{\rho_N }{4M_Nf_{D^*/D_1}^2M_{D^*/D_1}^3}a\, ;
\end{eqnarray}
the second term denotes  the single-pole term,
and corresponds to the off-shell (i.e. $\omega^2\neq M_{D^*/D_1}^2$) effects of the $T$-matrixes; and the
third term denotes   the continuum term or the  remaining effects,
where the $s_{0}$ is the continuum threshold.

In the  limit $\omega\rightarrow 0$, the
$T_{N}(\omega,{\bf 0})$ is equivalent to the Born term $T_{D^*/D_1N}^{\rm
Born}(\omega,{\bf 0})$. We take
into account the Born term at the phenomenological side,
\begin{eqnarray}
T_{N}(\omega^2)&=&T_{D^*/D_1N}^{\rm
Born}(\omega^2)+\frac{a}{(M_{D^*/D_1}^2-\omega^2)^2}+\frac{b}{M_{D^*/D_1}^2-\omega^2}+\frac{c}{s_0-\omega^2}\, ,
\end{eqnarray}
with the constraint
\begin{eqnarray}
\frac{a}{M_{D^*/D_1}^4}+\frac{b}{M_{D^*/D_1}^2}+\frac{c}{s_0}&=&0 \, .
\end{eqnarray}

The contributions from the intermediate spin-$\frac{3}{2}$ charmed
baryons are zero in the soft-limit $q_\mu \to 0$
\cite{Wangzg}, where the $q_\mu$ denotes the four-momentum   of the pseudoscalar mesons $P$  and vector mesons $V$ in the vertexes $B_{\frac{3}{2}}B_{\frac{1}{2}}P$ and $B_{\frac{3}{2}}B_{\frac{1}{2}}V$ for the spin-$\frac{3}{2}$ baryons $B_{\frac{3}{2}}$, spin-$\frac{1}{2}$ baryons  $B_{\frac{1}{2}}$ and mesons. The  contributions from the spin-$\frac{3}{2}$ charmed baryons, which are taken as the higher resonances,  are included in the full $D^*N\to D^*N$ scattering amplitude, see the paragraph after Eq.(15).  We take into account the intermediate
spin-$\frac{1}{2}$ charmed baryons in calculating the Born terms, and parameterize
 the hadronic matrix elements as
 \begin{eqnarray}
 \langle\Lambda_c/\Sigma_c(p-q)|D^*(-q)N(p)\rangle &=&\bar{U}_{\Lambda_c/\Sigma_c}(p-q)\left[ g_{\Lambda_c/\Sigma_cD^*N}\!\not\!{\epsilon}+i\frac{g^T_{\Lambda_c/\Sigma_cD^*N}}{M_N+M_{\Lambda_c/\Sigma_c}}\sigma^{\alpha\beta}\epsilon_\alpha q_\beta\right]U_N(p)\,,\nonumber\\
 \langle\Lambda_c/\Sigma_c(p-q)|D_1(-q)N(p)\rangle &=&\bar{U}_{\Lambda_c/\Sigma_c}(p-q)\left[ g_{\Lambda_c/\Sigma_cD_1N}\!\not\!{\epsilon}+i\frac{g^T_{\Lambda_c/\Sigma_cD_1N}}{M_N+M_{\Lambda_c/\Sigma_c}}\sigma^{\alpha\beta}\epsilon_\alpha q_\beta\right]\gamma_5U_N(p)\,,\nonumber\\
 \end{eqnarray}
 where the $U_N$ and $\bar{U}_{\Lambda_c/\Sigma_c}$ are the Dirac spinors of the nucleon and the charmed  baryons $\Lambda_c/\Sigma_c$, respectively;
 the $g_{\Lambda_c/\Sigma_cD^*N}$, $g_{\Lambda_c/\Sigma_cD_1N}$, $g_{\Lambda_c/\Sigma_cD^*N}^T$ and  $g_{\Lambda_c/\Sigma_cD_1N}^T$ are the strong  coupling constants in the vertexes. In the limit $q_\mu \to 0$, the  strong coupling constants $g_{\Lambda_c/\Sigma_cD^*N}^T$ and  $g_{\Lambda_c/\Sigma_cD_1N}^T$ have no contributions.

 We draw the Feynman  diagrams,  calculate the Born terms and obtain the results
\begin{eqnarray}
T_{D^*N}^{\rm Born}(\omega,{\bf0})&=&\frac{2f_{D^*}^2M_{D^*}^2M_N(M_H+M_N)g_{HD^*N}^2}
{\left[\omega^2-(M_H+M_N)^2\right]\left[\omega^2-M_{D^*}^2\right]^2}\,, \nonumber \\
T_{D_1N}^{\rm Born}(\omega,{\bf0})&=&\frac{2f_{D_1}^2M_{D_1}^2M_N(M_H-M_N)g_{HD_1N}^2}
{\left[\omega^2-(M_H-M_N)^2\right]\left[\omega^2-M_{D_1}^2\right]^2}\,,
\end{eqnarray}
where the  $H$ means either
$\Lambda_c^+$, $\Sigma_c^+$, $\Sigma_c^{++}$ or $\Sigma_c^0$. The masses $M_{\Lambda_c}=2.286\,\rm{GeV}$   and
$M_{\Sigma_c}=2.454\,\rm{GeV}$ from the Particle Data Group \cite{PDG}, we can take $M_H\approx
2.4\,\rm{GeV}$ as the average value.
On the other hand, there are no inelastic channels for the
$\bar{D}^*  N$ and $\bar{D}_1 N$ interactions in the case of the charmed mesons $\bar{c}q$.

  The scattering state $D^*N$  can translate  to the scattering states $D^*N$, $\pi\Sigma_c$, $\eta\Lambda_c$, $DN$, $\pi\Lambda_c$, $\rho\Sigma_c$, $\rho\Lambda_c$, etc, we can take into account the infinite series of the intermediate baryon-meson loops with the Bethe-Salpeter equation to obtain the full $D^*N \to D^*N$ scattering amplitude, and the higher resonances,  such as the $\Lambda_c(2595)$, $\Sigma_c(2800)$, etc,  appear as dynamically generated baryon states \cite{DN-BSE}. We can saturate the full $D^*N\to D^*N$ scattering amplitude with the tree-level Feynman diagrams of the  exchanges of the higher resonances, which have  energy dependent widths in the Breit-Wigner formulae. There are both spin-$\frac{1}{2}$ and spin-$\frac{3}{2}$ higher resonances, neglecting the spin-$\frac{3}{2}$ contributions can lead to  unknown uncertainties as the spin-$\frac{3}{2}$ contributions exist for $q_\mu\neq0$.
   The spin-$\frac{1}{2}$ higher resonances consist of  the negative-parity  charmed baryons $\Lambda_c(2595)$ and $\Sigma_c({\frac{1}{2}}^-)$ have the average mass $M_{H'}\approx 2.7\,\rm{GeV}$ \cite{PDG,Wang-Negative} and other excited  charmed baryons have even larger masses.
      The translations of the  scattering state $D^*N $ to the ground  states $\Lambda_c $ and $\Sigma_c$ are greatly facilitated in the phase-space.
    If the  couplings of the $D^*N$ to the spin-$\frac{1}{2}$ ground states and spin-$\frac{1}{2}$ higher resonances are of the same magnitude, we can neglect the higher resonances without impairing  the prediction ability  remarkably.  We admit that the imaginary parts of the inelastic scattering amplitudes therefore the imaginary parts of the scattering lengths  are lost by neglecting the loop-effects (or higher resonance contributions). The real and imaginary parts of the scattering amplitudes relate with the mass and width modifications in the nuclear matter, respectively, we expect that neglecting the imaginary part of the scattering amplitude cannot impair the prediction ability  remarkably (or  qualitatively) for the mass-shift $\delta M_{D^*}$. In calculations, we observe that the mass-shift $\delta M_{D^*}$ does not sensitive to contributions of the  ground  states $\Lambda_c $ and $\Sigma_c$,    see Table 1, the contributions from the spin-$\frac{1}{2}$ higher resonances maybe even smaller.     If we approximate the elastic scattering amplitudes plus the real parts of the full inelastic scattering amplitudes with the elastic scattering amplitudes at the phenomenological side, where the full inelastic scattering amplitudes contain contributions from the spin-$\frac{1}{2}$ and spin-$\frac{3}{2}$ higher resonances,  the mass-shifts obtained from the QCD sum rules appear as collective effects and receive  contributions from the real parts
    (not imaginary parts) of the inelastic scattering amplitudes.
In this article, we neglect the contributions from  the scattering states  (or continuum states) $\pi\Sigma_c$, $\eta\Lambda_c$, $DN$, $\pi\Lambda_c$, $\rho\Sigma_c$, $\rho\Lambda_c$, etc, which can lead to unknown uncertainties, but we still expect that  the prediction ability survives at least qualitatively.

 We carry out the operator
product expansion to the  condensates  up to dimension-5  at the large space-like  region in the nuclear matter,
and obtain the analytical expressions of the correlation functions at the level of quark-gluon degree's of  freedom,
 \begin{eqnarray}
\Pi_{\mu\nu}(q_0,\vec{q})&=&\left(-g_{\mu\nu}+\frac{q_\mu q_\nu}{q^2} \right)\sum_n C_n(q_0,\vec{q})\langle{\cal{O}}_n\rangle_{\rho_N}+\cdots\,,
\end{eqnarray}
where the $C_n(q_0,\vec{q})$ are the Wilson coefficients,  the in-medium condensates  $\langle{\cal{O}}_n\rangle_{\rho_N}  =\langle{\cal{O}}_n\rangle+\frac{\rho_N}{2M_N}\langle
{\cal{O}}_n\rangle_N$  at the low nuclear density, the   $\langle{\cal{O}}_n\rangle$ and $\langle
{\cal{O}}_n\rangle_N$ denote the vacuum condensates and nuclear matter induced condensates,  respectively. One can consult Refs.\cite{C-parameter,Drukarev1991} for the technical details in  the operator product expansion. Then we collect the terms proportional to $\rho_N$ (or the nuclear matter induced condensates),
    take the
quark-hadron duality,
 \begin{eqnarray}
T^N_{\mu\nu}(\omega,\vec{q})&=&\left(-g_{\mu\nu}+\frac{q_\mu q_\nu}{q^2} \right)\sum_n C_n(\omega,\vec{q})\langle{\cal{O}}_n\rangle_{N}+\cdots\, ,
\end{eqnarray}
 set  $\omega^2=q^2$, and perform the Borel transform  with respect
to the variable $Q^2=-\omega^2$, finally   obtain  the following two QCD  sum
rules:
\begin{eqnarray}
&& a \left\{\frac{1}{M^2}e^{-\frac{M_{D^*}^2}{M^2}}-\frac{s_0}{M_{D^*}^4}e^{-\frac{s_0}{M^2}}\right\}
+b \left\{e^{-\frac{M_{D^*}^2}{M^2}}-\frac{s_0}{M_{D^*}^2}e^{-\frac{s_0}{M^2}}\right\}
+ \frac{2f_{D^*}^2M_{D^*}^2M_N(M_H+M_N)g_{HD^*N}^2}{(M_H+M_N)^2-M_{D^*}^2}\nonumber\\
&&\left\{ \left[\frac{1}{(M_H+M_N)^2-M_{D^*}^2}-\frac{1}{M^2}\right]
e^{-\frac{M_{D^*}^2}{M^2}}-\frac{1}{(M_H+M_N)^2-M_{D^*}^2}e^{-\frac{(M_H+M_N)^2}{M^2}}\right\}=\left\{-\frac{m_c\langle\bar{q}q\rangle_N}{2}\right.\nonumber\\
&&\left.-\frac{2\langle q^\dag i D_0q\rangle_N}{3}+\frac{m_c^2\langle q^\dag i D_0q\rangle_N}{M^2}+\frac{m_c\langle\bar{q}g_s\sigma Gq\rangle_N}{3M^2}+\frac{8m_c\langle \bar{q} i D_0 i D_0q\rangle_N}{3M^2}-\frac{m_c^3\langle \bar{q} i D_0 i D_0q\rangle_N}{M^4}\right\}e^{-\frac{m_c^2}{M^2}}\nonumber\\
&&-\frac{1}{24}\langle\frac{\alpha_sGG}{\pi}\rangle_N\int_0^1dx \left(1+\frac{\widetilde{m}_c^2}{2M^2}\right)e^{-\frac{\widetilde{m}_c^2}{M^2}}
+\frac{1}{48M^2}\langle\frac{\alpha_sGG}{\pi}\rangle_N\int_0^1\frac{1-x}{x}\left(\widetilde{m}_c^2-\frac{\widetilde{m}_c^4}{M^2}\right)e^{-\frac{\widetilde{m}_c^2}{M^2}}\, ,
\end{eqnarray}

\begin{eqnarray}
&& a \left\{\frac{1}{M^2}e^{-\frac{M_{D_1}^2}{M^2}}-\frac{s_0}{M_{D_1}^4}e^{-\frac{s_0}{M^2}}\right\}
+b \left\{e^{-\frac{M_{D_1}^2}{M^2}}-\frac{s_0}{M_{D_1}^2}e^{-\frac{s_0}{M^2}}\right\}
+ \frac{2f_{D_1}^2M_{D_1}^2M_N(M_H-M_N)g_{HD_1N}^2}{(M_H-M_N)^2-M_{D_1}^2}\nonumber\\
&&\left\{ \left[\frac{1}{(M_H-M_N)^2-M_{D_1}^2}-\frac{1}{M^2}\right]
e^{-\frac{M_{D_1}^2}{M^2}}-\frac{1}{(M_H-M_N)^2-M_{D_1}^2}e^{-\frac{(M_H-M_N)^2}{M^2}}\right\}=\left\{\frac{m_c\langle\bar{q}q\rangle_N}{2}\right.\nonumber\\
&&\left.-\frac{2\langle q^\dag i D_0q\rangle_N}{3}+\frac{m_c^2\langle q^\dag i D_0q\rangle_N}{M^2}-\frac{m_c\langle\bar{q}g_s\sigma Gq\rangle_N}{3M^2}-\frac{8m_c\langle \bar{q} i D_0 i D_0q\rangle_N}{3M^2}+\frac{m_c^3\langle \bar{q} i D_0 i D_0q\rangle_N}{M^4}\right\}e^{-\frac{m_c^2}{M^2}}\nonumber\\
&&-\frac{1}{24}\langle\frac{\alpha_sGG}{\pi}\rangle_N\int_0^1dx \left(1+\frac{\widetilde{m}_c^2}{2M^2}\right)e^{-\frac{\widetilde{m}_c^2}{M^2}}
+\frac{1}{48M^2}\langle\frac{\alpha_sGG}{\pi}\rangle_N\int_0^1\frac{1-x}{x}\left(\widetilde{m}_c^2-\frac{\widetilde{m}_c^4}{M^2}\right)e^{-\frac{\widetilde{m}_c^2}{M^2}}\, ,
\end{eqnarray}
where $\widetilde{m}_c^2=\frac{m_c^2}{x}$.

Differentiate  above equation with respect to  $\frac{1}{M^2}$, then
eliminate the
 parameter $b$, we can obtain the QCD sum rules for
 the parameter $a$. With the simple replacements $m_c \to m_b$, $M_{D^*} \to M_{B^*}$, $M_{D_1} \to M_{B_1}$, $\Lambda_c \to \Lambda_b$ and $\Sigma_c \to \Sigma_b$,
  we can obtain the corresponding QCD sum rules for
 the mass-shifts of the $B^*$ and $B_1$ mesons in the nuclear matter, where we take the approximation $M_H=\frac{M_{\Sigma_b}+M_{\Lambda_b}}{2}\approx 5.7\,\rm{GeV}$ \cite{PDG}.

The present approach was introduced by  Koike and Hayashigaki to study   the spin-isospin averaged meson-nucleon scattering lengths and
the relevant mass-shifts for the  $\rho$, $\omega$, $\phi$ mesons in the  nuclear matter \cite{Koike1996}. The heavy mesons
 contain  a  heavy-quark and a light-quark.
The existence of a heavy quark in the heavy mesons results in much
difference between the in-medium properties of
 the heavy mesons and light mesons.
The heavy quark  interacts  with the nuclear matter through  the exchange of the intermediate gluons   and the modifications of the gluon condensates in the nuclear matter are mild, while the modifications of the quark condensates in the nuclear matter are rather large.   We expect that the convergent behaviors  of the heavy-light type interpolating   currents are better than that of the light-light type interpolating currents, if  the correlation functions $\Pi_{\mu\nu}(q)$ are expanded in terms of the external parameter $\rho_N$. The
 approach developed for the light mesons still works for the heavy mesons in the operator product expansion side. In the phenomenological side, we take the lowest order Born terms plus the elastic scattering amplitudes to approximate the phenomenological spectrum,   the present article shares both the advantages and shortcomings of the approach developed in Ref.\cite{Koike1996}.

In Ref.\cite{Klingl-Refe},   Klingl, Kaiser and Weise use
   an effective Lagrangian which combines chiral $SU(3)$ dynamics with
vector meson dominance to calculate the forward vector-meson-nucleon
scattering amplitudes, and take them as input parameters in the hadronic side of the QCD sum rules in nuclear matter, and
observe a remarkable degree of consistency with the operator product expansion at the quark level.
In Ref.\cite{Leupold-Refe},   Leupold and Mosel
  study the electromagnetic current-current correlation functions in the nuclear matter, and expand the QCD sum rules
   in terms of the finite squared three-momentum $\vec{q}^2$, and observe that the QCD sum rule can provide  an interesting and non-trivial
consistency check for the hadronic models, but cannot rule out the hadronic
models which predict a different behavior of the vector mesons with different $\vec{q}^2$.  In this article, the hadronic side of the QCD sum rules of the
order ${\cal O }(\rho_N)$ consists of the elastic scattering amplitudes  and the lowest Born terms,  and can lead to stable QCD sum rules with the suitable Borel parameters in a finite range, so the hadronic model is consistent with the operator product expansion side   with $\vec{q}^2=0$ \cite{Leupold-Refe}.
 According to Ref.\cite{Klingl-Refe}, there are both real and imaginary parts in the  scattering amplitudes, neglecting the imaginary parts of the  scattering amplitudes therefore the imaginary parts of the  scattering lengths miss the inelastic contributions such as the transitions $D^*N \to$, $\pi\Sigma_c$, $\eta\Lambda_c$, $DN$, $\pi\Lambda_c$, $\rho\Sigma_c$, $\rho\Lambda_c$, etc, the approximation results in unwanted  uncertainties, it is the shortcoming of the present method. We can calculate the scattering amplitudes with the effective field theory based on the heavy quark symmetry
and chiral symmetry, and keep the imaginary parts of the scattering amplitudes explicitly, and study the loop effects, the tedious calculation is beyond the present work.

\section{Numerical results and discussions}
In calculations, we have assumed that  the linear density
approximation   is valid at the low nuclear  density,
$\langle{\cal{O}}\rangle_{\rho_N}=\langle0|{\cal{O}}|0\rangle+\frac{\rho_N}{2M_N}\langle
N|{\cal{O}}|N\rangle=\langle{\cal{O}}\rangle+\frac{\rho_N}{2M_N}\langle
{\cal{O}}\rangle_N$ for a general
condensate $\langle{\cal{O}}\rangle_{\rho_N}$ in the nuclear matter. The input parameters  are taken as
  $\langle\bar{q} q\rangle_N={\sigma_N \over m_u+m_d } (2M_N)$,
 $\langle\frac{\alpha_sGG}{\pi}\rangle_N= (-0.65\pm 0.15) \,{\rm {GeV}} (2M_N)$,
$\langle q^\dagger iD_0 q\rangle_N=(0.18\pm0.01) \,{\rm{GeV}}(2M_N)$,
$\langle\bar{q}g_s\sigma G q\rangle=3.0\,{\rm GeV}^2(2M_N) $,
$\langle \bar{q} iD_0iD_0
q\rangle_N+{1\over8}\langle\bar{q}g_s\sigma G
q\rangle_N=0.3\,{\rm{GeV}}^2(2M_N)$, $m_u+m_d=12\,\rm{MeV}$,
$\sigma_N=(45\pm 10)\,\rm{MeV}$, $M_N=0.94\,\rm{GeV}$,
$\rho_N=(0.11\,\rm{GeV})^3$ \cite{C-parameter}, $m_c=(1.35\pm 0.1)\,\rm{GeV}$ and $m_b=(4.7\pm 0.1)\,\rm{GeV}$ at the energy scale $\mu=1\,\rm{GeV}$.

The value of the strong coupling constant $g_{DN\Lambda_c}$ is $g_{\Lambda_c DN}=6.74$ from
the  QCD sum rules \cite{Nielsen98}, while the average value of the strong coupling constants $g_{\Lambda_cDN}$ and $g_{\Sigma_cDN}$ from the light-cone QCD sum rules is $\frac{g_{\Lambda_cDN}+g_{\Sigma_cDN}}{2}=6.775$ \cite{Khodjamirian1108}, those values are consistent with each other.
The average value of the strong coupling constants $g_{\Lambda_c D^*N}$ and $g_{\Sigma_c D^*N}$ from the light-cone QCD sum rules is  $\frac{g_{\Lambda_c D^*N}+g_{\Sigma_cD^*N}}{2}=3.86$ \cite{Khodjamirian1108}.  In this
article, we take the approximation $g_{\Lambda_cD^*N}\approx g_{\Sigma_cD^*N}
\approx g_{\Lambda_cD_1N}\approx g_{\Sigma_cD_1N}\approx g_{\Lambda_bB^*N}\approx g_{\Sigma_bB^*N}
\approx g_{\Lambda_bB_1N}\approx g_{\Sigma_bB_1N}\approx3.86$.

For the well established vector mesons $D^*$ and $B^*$, we take the values from the Particle Data Group,
$M_{D^*}=2.01\,\rm{GeV}$  and $M_{B^*}=5.325\,\rm{GeV}$ \cite{PDG},  the decay constants  $f_{D^*}$ and $f_{B^*}$
are determined by  the QCD sum rules,   $f_{D^*}=0.270\,\rm{GeV}$ and
$f_{B^*}=0.195\,\rm{GeV}$ \cite{SRreview}, where the threshold parameters are taken as
$s^0_{D^*}=(5-7)\,\rm{GeV}^2$  and $s^0_{B^*}=(33-37)\,\rm{GeV}^2$ \cite{SRreview,Kho-Decay}, here we have neglected the uncertainties of the
decay constants.
We can take the threshold parameters   as
$s^0_{D^*}=6.5\,\rm{GeV}^2$, $s^0_{B^*}=35\,\rm{GeV}^2$ and the Borel parameters as $T^2_{D^*}=(1.6-2.6)\,\rm{GeV}^2$, $T^2_{B^*}=(4.0-6.0)\,\rm{GeV}^2$ to reproduce the values
$M_{D^*}=2.01\,\rm{GeV}$, $M_{B^*}=5.325\,\rm{GeV}$, $f_{D^*}=0.270\,\rm{GeV}$ and
$f_{B^*}=0.195\,\rm{GeV}$ approximately for the QCD sum rules in the vacuum \footnote{We use the $T^2$ to denote the Borel parameters in the vacuum.}.
The mass of the axialvector meson $D^{0}_1(2430)$ is
$M_{D_1}=(2427\pm26\pm25)\,\rm{MeV}$ from the Particle Data Group \cite{PDG}, and the axialvector meson $B_1$ has  not been observed yet.
   We calculate the hadronic parameters of the axialvector mesons $D_1$ and $B_1$
using the QCD sum rules in the vacuum, and obtain the values  $M_{D_1}=2.42\,\rm{GeV}$,
$M_{B_1}=5.75\,\rm{}GeV$, $f_{D_1}=0.305\,\rm{GeV}$ and
$f_{B_1}=0.255\,\rm{GeV}$ with the threshold parameters $s^0_{D_1}=8.5\,\rm{GeV}^2$, $s^0_{B_1}=39\,\rm{GeV}^2$ and the Borel parameters $T^2_{D_1}=(2.0-3.0)\,\rm{GeV}^2$, $T^2_{B_1}=(5.0-7.0)\,\rm{GeV}^2$.
 The value $M_{D_1}=2.42\,\rm{GeV}$ reproduces the experimental data $M_{D_1}=(2427\pm26\pm25)\,\rm{MeV}$ well \cite{PDG}. The prediction of the mass
  $M_{B_1}$ satisfies the relation $M_{B_1}-M_{B^*}\approx M_{D_1}-M_{D^*}$.
For the explicit expressions of the QCD sum rules in the vacuum derived from the correlation functions $\Pi^{0}_{\mu\nu}(q)$, one can consult Ref.\cite{Wang-CPL} and the references therein.

In the QCD sum rules, the phenomenological  hadronic spectrum does not depend on the Borel parameters, the continuum threshold parameters $\sqrt{s_0}$ are usually  taken as $\sqrt{s_0}=M_{gr}\sim M_{ra}$, where the $gr$ and $ra$ denote the ground states and first radial excited states respectively, so there are  uncertainties come from the continuum threshold parameters. We choose suitable Borel parameters to satisfy the two criteria (pole dominance and convergence of the operator product
expansion) of the QCD sum rules. The optimal Borel parameters result in rather stable  QCD sum rules which are not sensitive to the  threshold parameters and have
 Borel platforms. The  phenomenological  hadronic spectrum in a definite channel (in other words, for a definite interpolating current) survives in different QCD sum rules, we can choose the same threshold parameters but different Borel parameters to satisfy the two  criteria of the QCD sum rules in studying the correlation functions $\Pi^0_{\mu\nu}$ and $T_{\mu\nu}^N$.  In this article, the threshold parameters are taken as
$s^0_{D^*}=(6.5\pm0.5)\,\rm{GeV}^2$, $s^0_{D_1}=(8.5\pm0.5)\,\rm{GeV}^2$, $s^0_{B^*}=(35\pm1)\,\rm{GeV}^2$ and $s^0_{B_1}=(39\pm1)\,\rm{GeV}^2$,
respectively, which   satisfy the relations $s^0_{D^*,D_1,B_1}=(M_{D^*,D_1,B_1}+0.4\sim 0.6\, \rm{GeV})^2$ and $s^0_{B^*}=(M_{B^*}+0.5\sim 0.7\, \rm{GeV})^2$.
In general, the energy gap between the ground state and the first radial excited state is about $0.5\,\rm{GeV}$.

In Fig.1, we plot the mass-shifts  $\delta M$ versus the Borel
parameter $M^2$ at large intervals. From the figure, we can see that the values of the
mass-shifts  are rather stable with variations  of the Borel
parameter at the intervals $M^2=(4.5-5.4)\,\rm{GeV}^2$, $(6.5-7.6)\,\rm{GeV}^2$,
$(22-24)\,\rm{GeV}^2$ and $(34-37)\,\rm{GeV}^2$ for the $D^*$, $D_1$, $B^*$ and $B_1$ mesons,
respectively; in other words, the uncertainties originate from the Borel parameter
$M^2$ are less than $1\%$.   The main contributions come from the terms $\pm m_c\langle\bar{q}q\rangle_N$ and   $\pm m_b\langle\bar{q}q\rangle_N$, see Eqs.(18-19) and Fig.2, the operator product expansion is well convergent.
  The  spectral densities at the level of the quark-gluon degrees of freedom consist of the medium-induced condensates, and have the form $A_1 \delta(s-m_{c/b}^2)+A_2\delta(s-\widetilde{m}_{c/b}^2)$, where the $A_1$ and $A_2$ denote the coefficients, we carry out the integrals,
  \begin{eqnarray}
  \int_{m_{c/b}^2}^{s_0}ds\left[ A_1 \delta(s-m_{c/b}^2)+A_2\delta(s-\widetilde{m}_{c/b}^2)\right]e^{-\frac{s}{M^2}}\, ,
  \end{eqnarray}
  to obtain the right side of the QCD sum rules in Eqs.(18-19), there are no perturbative terms  to approximate the continuum contributions at the regions $s>s_0$.
 At the phenomenological side,  the  exponential  factors
 \begin{eqnarray}
 e^{-\frac{s_0}{M^2}}&=&e^{-(1.20-1.44)}, \, e^{-(1.12-1.31)}, \,e^{-(1.46-1.59)},  \,e^{-(1.05-1.15)},
 \end{eqnarray}
 at the intervals $M^2=(4.5-5.4)\,\rm{GeV}^2$, $(6.5-7.6)\,\rm{GeV}^2$,
$(22-24)\,\rm{GeV}^2$ and $(34-37)\,\rm{GeV}^2$ for the $D^*$, $D_1$, $B^*$,  $B_1$ mesons
respectively,  where  we take the central values of the threshold parameters, the corresponding  exponential  factors  of the ground states are
\begin{eqnarray}
  e^{-\frac{M^2_m}{M^2}}&=&e^{-(0.75-0.90)}, \,e^{-(0.77-0.90)}, \,e^{-(1.18-1.29)}, \,e^{-(0.89-0.97)},
 \end{eqnarray}
where the $m$ stands for the $D^*$, $D_1$, $B^*$,  $B_1$ mesons respectively; the continuum contributions are suppressed  more efficiently.
 Furthermore, we expect that the couplings of
 an special interpolating current to the excited states are more weak than that to the ground state mesons. For example, the decay constants of the pseudoscalar mesons $\pi(140)$ and $\pi(1800)$ have the hierarchy: $f_{\pi(1300)}\ll f_{\pi(140)}$ from the Dyson-Schwinger equation \cite{CDRoberts}, the lattice QCD \cite{Latt-pion},  the QCD sum rules \cite{QCDSR-pion}, etc, or from the experimental data \cite{pion-exp}.

We can take  the Borel windows as $M^2=(4.5-5.4)\,\rm{GeV}^2$, $(6.5-7.6)\,\rm{GeV}^2$,
$(22-24)\,\rm{GeV}^2$ and $(34-37)\,\rm{GeV}^2$ for the $D^*$, $D_1$, $B^*$ and $B_1$ mesons,
respectively, and obtain mass-shifts
$\delta M_{D^*}=-71^{+20}_{-23}\,\rm{MeV}$, $\delta M_{B^*}=-380^{+82}_{-91}\,\rm{MeV}$, $\delta M_{D_1}=72^{+22}_{-20}\,\rm{MeV}$,  $\delta M_{B
_1}=264^{+76}_{-69}\,\rm{MeV}$, respectively; and the scattering lengths
 $a_{D^*}=-1.07^{+0.30}_{-0.34}\,\rm{fm}$, $a_{B^*}=-7.17^{+1.53}_{-1.71}\,\rm{fm}$,  $a_{D_1}=1.15^{+0.35}_{-0.32}\,\rm{fm}$ and $a_{B_1}=5.03^{+1.46}_{-1.31}\,\rm{fm}$ for the $D^*N$, $B^*N$, $D_1N$ and $B_1N$ interactions, respectively. For the technical details in analyzing the uncertainties, one can consult Ref.\cite{Baxi}.

In Fig.3, we plot the  mass-shifts $\delta M$ versus the Borel
parameter $M^2$ and the strong coupling constants $g^2$. From the figure, we can see that the mass-shifts decrease (increase)
 monotonously with increase of the squared strong coupling constants $g^2$ for the vector mesons $D^*$ and $B^*$ (axialvector mesons $D_1$ and $B_1$)
 in the Borel windows. The precise values of the mass-shifts and scattering lengths are presented in Table 1.

Although the present QCD sum rules are not stable with variations of the Borel parameters, the uncertainties originate  from the Borel parameters are less than $1\%$ in the Borel windows, i.e. we  choose suitable platforms to avoid large uncertainties. On the other hand, we can take moments of the correlation functions and derive QCD sum rules to study the mass-shifts as in Ref.\cite{DB-SR},
the present QCD sum rules are equal to that derived from the first two moments of the correlation functions, i.e. we take the weight functions to be $1$ and $s$.

In the present work and Refs.\cite{Hayashigaki,WangHuang},  the correlation functions  are  divided
into  a vacuum part   and a static one-nucleon part, and  the nuclear matter  induced effects are extracted  explicitly;
  while in Refs.\cite{DB-SR,SumRule-D0},   the pole terms of the     hadronic spectral densities  are  parameterized
as $\frac{\rm{Im}\Pi(\omega,0)}{\pi}=F_{+}\delta(\omega-M_{+})-F_{-}\delta(\omega+M_{-})$, where $M_{\pm}=M\pm\Delta M$ and $F_{\pm}=F\pm\Delta F$,
and  QCD  sum rules for the   mass center $M$ and the mass splitting $\Delta M$ are obtained. For the pseudoscalar $D$, $\bar{D}$ mesons,
 Hayashigaki  obtains the mass-shift $\delta M_{D}=-50\,\rm{MeV}$ \cite{Hayashigaki},
while  Hilger, Thomas and Kampfer obtain the mass-shift
$\delta M_{D}=+45\,\rm{MeV}$ \cite{DB-SR}.
For scalar $D_0$, $\bar{D}_0$  mesons, the  mass-shift   $\delta M_{D_0}=M-M_{D_0}<0$   obtained by
Hilger and Kampfer \cite{SumRule-D0} differs  from the result $\delta M_{D0} = +69\,\rm{MeV}$ obtained by Wang and Huang \cite{WangHuang}.
 In Ref.\cite{SumRule-VA}, Hilger,   Kampfer and Leupold study the chiral partners of charmed mesons in the nuclear matter, and focus on the differences
 between the pseudoscalar and scalar as well as vector and axialvector D mesons and derive the corresponding
Weinberg type sum rules, while the mass-shifts are not presented. In the present work and Refs.\cite{Hayashigaki,WangHuang}, the inelastic scattering amplitudes (or the loop-effects) are neglected,  which can lead to unknown uncertainties; while in Refs.\cite{DB-SR,SumRule-D0}, the simple hadronic spectral densities  miss the complex energy dependence.
The two approaches both have shortcomings, and the corresponding predictions can be confronted with the experimental data in the future.

In the limit $m_q \to 0$, the  quark condensate $\langle\bar{q}q\rangle$ serves as the order parameter  and
indicates that the chiral symmetry is broken.
The quark condensate undergoes  reduction in the nuclear matter, $\langle\bar{q}q\rangle_{\rho_N}=\langle\bar{q}q\rangle+\frac{\rho_N}{2M_N}\langle\bar{q}q\rangle_N$,
    the chiral symmetry is partially  restored, for example, the in-medium nucleon mass $ M_N^*$  can be approximated as
 $M_N^*=-\frac{8\pi^2}{M^2}\langle\bar{q}q\rangle_{\rho_N}$ and the mass reduction is rather large;  on the other hand, there appear new medium-induced condensates, for example, the $\langle \bar{q} iD_0iD_0q\rangle_N$, $\langle\bar{q}g_s\sigma G q\rangle_N$, etc, which also break the chiral symmetry. In the present case, the medium-induced condensates  are associated with the large heavy quark masses $m_Q$, $m_Q^2$, $m_Q^3$ or $m_Q^4$, the net effects do not always warrant that  the chiral symmetry is monotonously restored with the increase of the density of the  nuclear matter. The light vector current $\bar{q}\gamma_\mu q$  and axialvector current $\bar{q}\gamma_\mu\gamma_5 q$ are invariant under the chiral transformation $q\to e^{i\alpha\gamma_5}q$, however, the heavy vector current $\bar{Q}\gamma_\mu q$  and axialvector current $\bar{Q}\gamma_\mu\gamma_5 q$ are mixed with each other under the transformation, the heavy quark currents $\bar{Q}\gamma_\mu q$  and  $\bar{Q}\gamma_\mu\gamma_5 q$ are not conserved in the limit $m_q \to 0$, it is better to take the doublets  $(D^*,D_1)$, $(B^*,B_1)$ as  the parity-doublets rather than  the chiral-doublets.
 If we take into account the flavor $SU(3)$ symmetry  of the light quarks, the chiral $SU(3)_L \times SU(3)_R$ transformations require that the ground states $(\bar{D}^{*0}, D^{*-}, D^{*-}_s)$ and $(B^{*+}, B^{*0}, B^{*0}_s)$ have their chiral partners  $(\bar{D}^{0}_1, D^{-}_1, D^{-}_{s1})$ and $(B^{+}_1, B^{0}_1, B^{0}_{s1})$, respectively, those parity-doublets are  chiral-doublets.
 When the density of the nuclear matter is large enough, the order parameter $\langle\bar{q}q\rangle_{\rho_N} \to 0$, the chiral symmetry is restored, the Fermi  gas  approximation for the nuclear matter does not survive, there are free of the non-perturbative contributions from the condensates, and the parity-doublets
 (or chiral-doublets) maybe have degenerated  masses approximately. In the present case, we study the parity-doublets (or chiral-doublets) in the low nuclear density, the mass breaking effects of the parity-doublets (or chiral-doublets)  maybe even larger.

The axialvector current $\bar{u}(x)\gamma_\mu \gamma_5 c(x)$ interpolates the axialvector meson $D_1(2430)$ has non-vanishing coupling
with the scattering state $D^*\pi$, in the soft $\pi$ limit, the  coupling constant $\lambda_{D^*\pi}$ can be estimated as
\begin{eqnarray}
\langle 0|\bar{u}(x)\gamma_\mu \gamma_5 c(x)|D^*\pi\rangle&=&-\frac{i}{f_\pi}\langle 0|\left[Q_5,\bar{u}(x)\gamma_\mu \gamma_5 c(x)\right]|D^*\rangle \nonumber \\
&=&-\frac{i}{f_\pi}\langle 0|\bar{d}(x)\gamma_\mu  c(x)|D^*\rangle =-\frac{if_{D^*}M_{D^*}\epsilon_\mu}{f_\pi}=i\lambda_{D^*\pi}\epsilon_\mu \, ,\\
&=&f_{D_1}M_{D_1}\varepsilon_\mu \frac{i}{p^2-M_{D_1}^2} \langle D_1|D^*\pi\rangle \, ,
\end{eqnarray}
where the axial-charge $Q_5=\int d^3y d^{\dagger}(y) \gamma_5 u(y)$, and the $\epsilon_\mu$ and $\varepsilon_\mu$ are
 the polarization  vectors  of the vector and axialvector mesons $D^*$ and $D_1$, respectively \cite{LeeDa}, the formula in Eq.(24) survives beyond the soft $\pi$ limit. The  coupling constant $\lambda_{D^*\pi}$  is a large quantity and  cannot be neglected. The rescatterings
 \begin{eqnarray}
 D^*\pi &\to& D^*\pi \, , \nonumber \\
 D^*\pi &\to& D^*\pi, D_s^*\bar{K}, D^* \eta \to D^*\pi\,,  \nonumber \\
 D^*\pi &\to& D^*\pi, D_s^*\bar{K}, D^* \eta \to D^*\pi, D_s^*\bar{K}, D^* \eta\to D^*\pi \, ,  \nonumber \\
 D^*\pi &\to& D^*\pi, D_s^*\bar{K}, D^* \eta \to D^*\pi, D_s^*\bar{K}, D^* \eta\to D^*\pi, D_s^*\bar{K}, D^* \eta\to D^*\pi \, ,
 \end{eqnarray}
$\cdots$ also have contributions to the hadronic spectral densities.
In the heavy meson chiral unitary approach, we can use the Bethe-Salpeter equation
to perform the summation of the infinite series of the intermediate meson-loops (such as the $D^*\pi$, $D_s^*\bar{K}$, $D^* \eta$) to obtain the full $D^*\pi \to D^*\pi$ scattering amplitude, and generate  the axialvector meson $D_1(2430)$ dynamically \cite{PANDA,GuoFK}. If we saturate the full  $D^*\pi \to D^*\pi$ scattering amplitude with the exchanges of the intermediate axialvector meson $D_1(2430)$,   the $D^*\pi$ rescattering effects lead to the renormalization  $\frac{1}{p^2-M_{D_1}^2+i\epsilon} \to \frac{1}{p^2-M_{D_1}^2-\widehat{\Pi}(p)+i\epsilon}$ in the hadronic representation of the correlation functions, where the $\widehat{\Pi}(p)$ denotes   the
renormalized self-energy   of the intermediate $D^*\pi$ loops, and contributes  a finite imaginary part to modify the dispersion relation.
In fact, the contributions of the intermediate meson-loops are very large, we have to take the mass $M_{D_1}$ as the bare mass
$\stackrel{ \rm \circ} M_{D_1}$ to absorb
the real part of  the un-renormalized self-energy to reproduce the physical mass, the net effects  are embodied in the finite imaginary part.
We can take
into account those meson-loops effectively  by taking the following
replacement for the hadronic spectral density,
\begin{eqnarray}
\delta\left(s-M_{D_1}^2\right) &\to& \frac{1}{\pi}\frac{\sqrt{s}\Gamma_{D_1}(s)}{\left(s-M_{D_1}^2\right)^2+s\Gamma_{D_1}^2(s)} \, ,
\end{eqnarray}
here we neglect the complicated renormalization procedure for simplicity \cite{WangIJTP}. Furthermore, we neglect the energy dependence of the width, and approximate it
with the experimental value from the Particle Data Group.
In Ref.\cite{WangIJTP}, we observe that a width  about (or less than) $400\,\rm{MeV}$ cannot
change the prediction significantly, the $\delta$ function approximation for the spectral densities still survives. In the present case,
 $\Gamma_{D_1}=384^{+107}_{-75}\pm 75\,\rm{MeV}$  from the Particle Data Group \cite{PDG}, the contaminations from the intermediate state $D^*\pi$ are expected
 to be small. Analogical discussions can be applied to the  contaminations from the intermediate state $B^*\pi$.

The negative scattering lengths $a_{D^*}=-1.07\,\rm{fm}$ and $a_{B^*}=-7.17\,\rm{fm}$ indicate   that
the $D^*N$ and $B^*N$ interactions are attractive, it is possible to
form the $D^*N$ and $B^*N$ bound states; while
the positive scattering lengths $a_{D_1}=1.15\,\rm{fm}$  and $a_{B_1}=5.03\,\rm{fm}$ indicate that the $D_1N$ and $B_1N$
interactions are repulsive, it is difficult  to form the $D_1N$ and
$B_1N$ bound states. According to the observations of the present work and Refs.\cite{Hayashigaki,WangHuang}, we can draw
the conclusion tentatively that the negative parity heavy mesons decrease their masses in the nuclear matter while the positive parity
heavy mesons increase their masses.   The decays of the high
charmonium states to the  negative parity charmed meson pairs $D\bar{D}$ and $D^*\bar{D}^*$ are facilitated, while the decays
to the positive parity  charmed mesons pairs $D_0\bar{D}_0$ and $D_1\bar{D}_1$ are suppressed.
 The $J/\psi$ production can obtain additional suppressions due to
 mass modifications of the negative parity charmed  mesons  $D$ and $D^*$
  in the  nuclear matter.

\begin{table}
\begin{center}
\begin{tabular}{|c|c|c|c|c|c|c|c|}\hline\hline
  $g^2$                    &0           &10         &20         &30         &40         &50 \\ \hline
  $\delta M_{D^*}$ (MeV)   &$-75$       &$-72$      &$-70$      &$-67$      &$-65$      &$-62$ \\ \hline
  $\delta M_{B^*}$ (MeV)   &$-382$      &$-381$     &$-380$     &$-380$     &$-379$     &$-378$ \\ \hline
  $\delta M_{D_1}$ (MeV)   &$70$        &$71$       &$73$       &$74$       &$76$       &$78$ \\ \hline
  $\delta M_{B_1}$ (MeV)   &$262$       &$263$      &$264$      &$265$      &$266$      &$267$ \\ \hline
  $a_{D^*}$ (fm)           &$-1.13$     &$-1.09$    &$-1.05$    &$-1.02$    &$-0.98$    &$-0.94$ \\ \hline
  $a_{B^*}$ (fm)           &$-7.20$     &$-7.18$    &$-7.17$    &$-7.15$    &$-7.14$    &$-7.13$ \\ \hline
  $a_{D_1}$ (fm)           &$1.11$      &$1.14$     &$1.16$     &$1.19$     &$1.21$     &$1.24$ \\ \hline
  $a_{B_1}$ (fm)           &$5.00$      &$5.02$     &$5.04$     &$5.05$     &$5.07$     &$5.09$ \\ \hline
\end{tabular}
\end{center}
\caption{ The
mass-shifts $\delta M$ and the scattering lengths $a$ versus the strong coupling constants $g^2$.  }
\end{table}

  \begin{figure}
  \centering
  \includegraphics[totalheight=6cm,width=7cm]{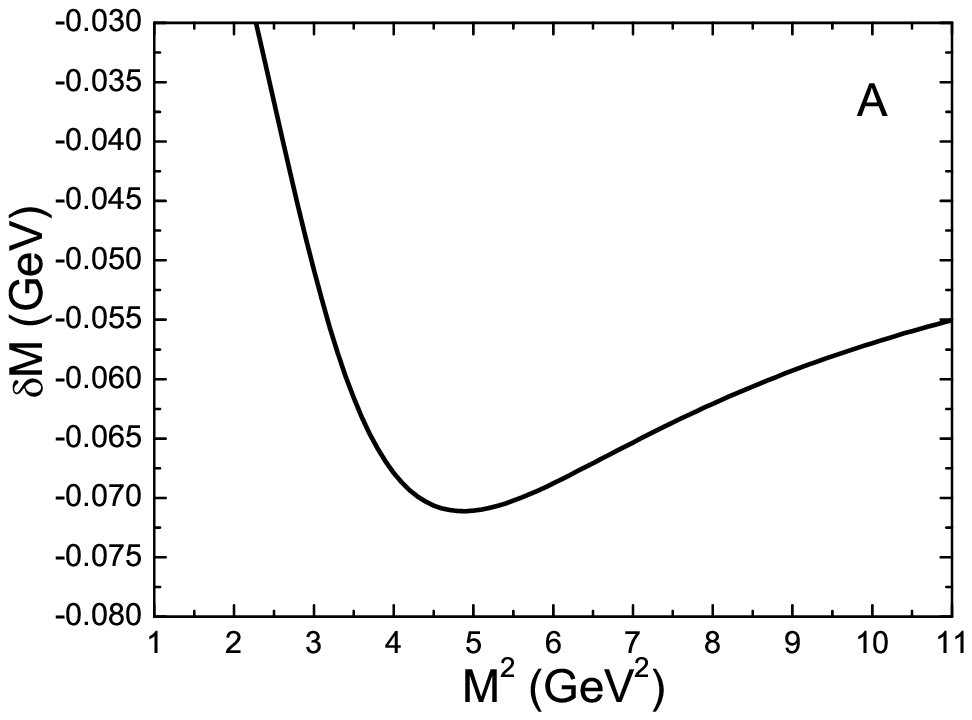}
  \includegraphics[totalheight=6cm,width=7cm]{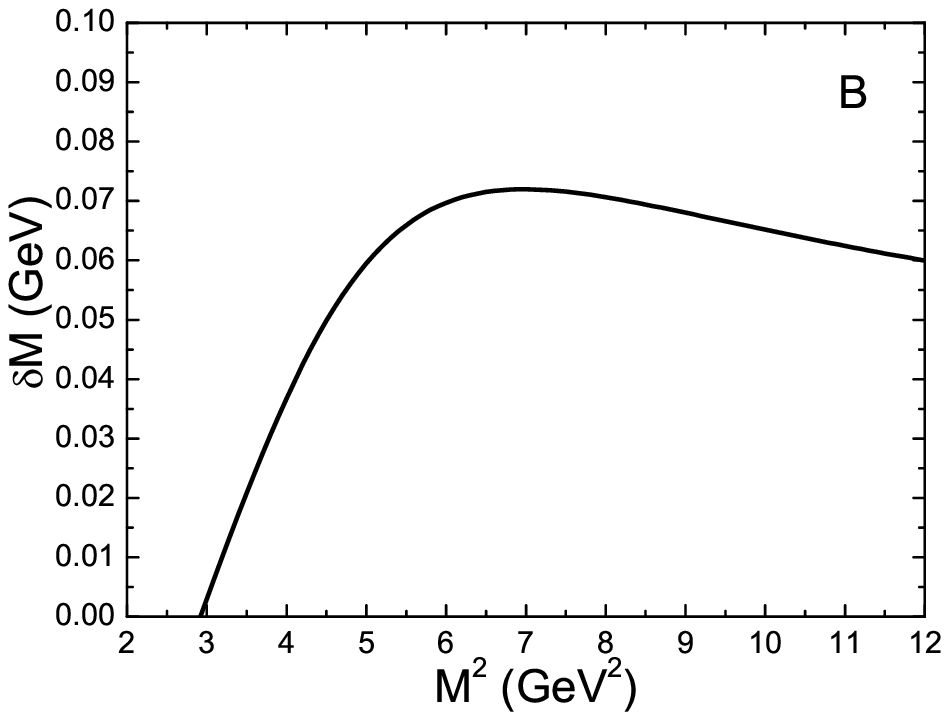}
   \includegraphics[totalheight=6cm,width=7cm]{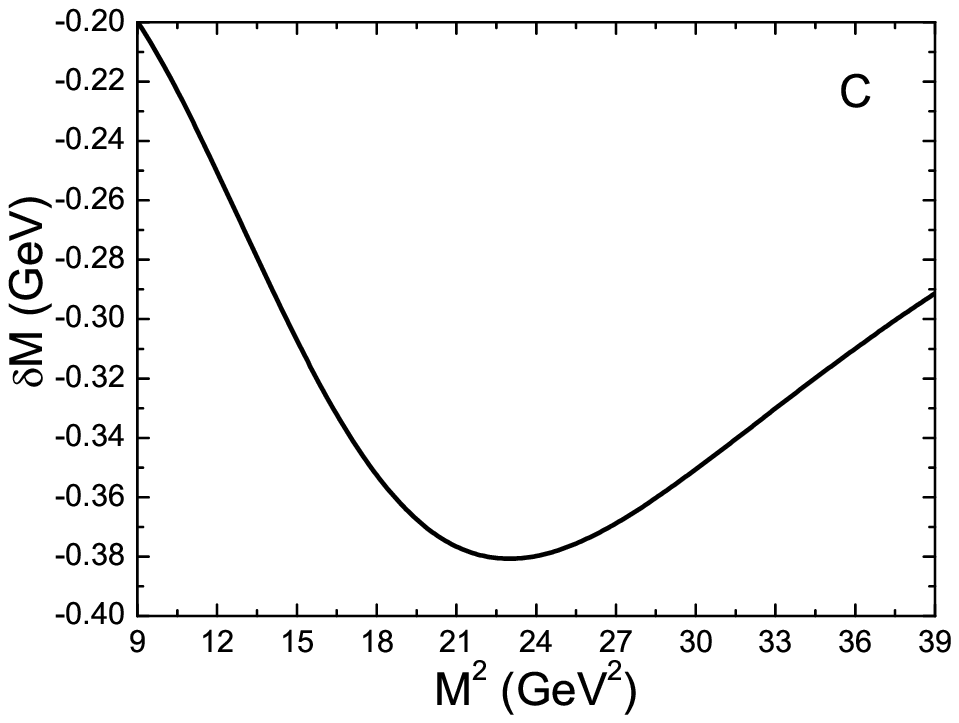}
    \includegraphics[totalheight=6cm,width=7cm]{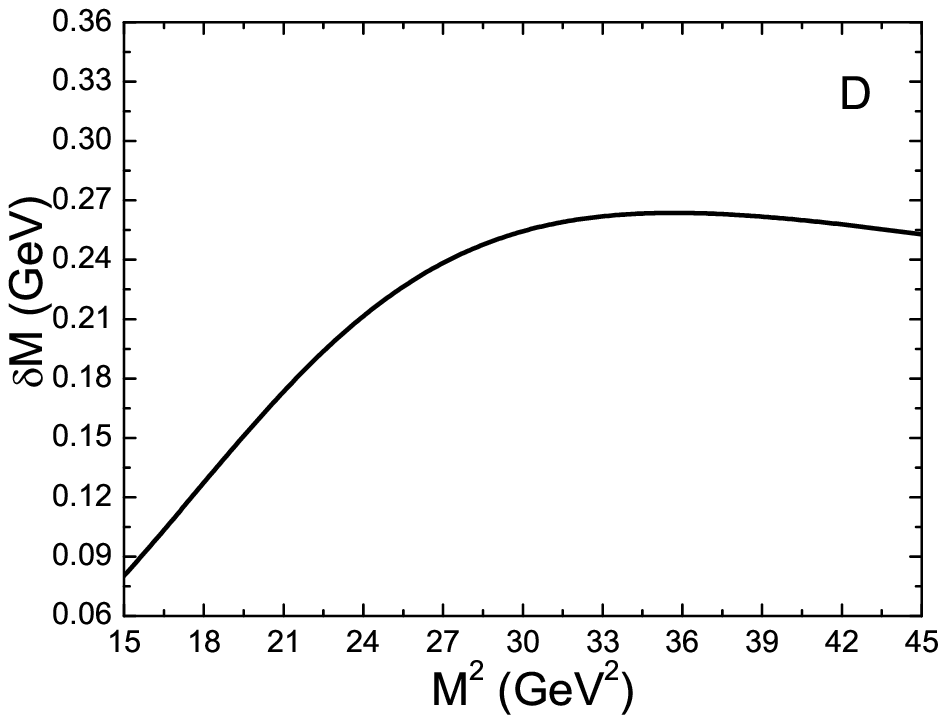}
  \caption{The   mass-shifts $\delta M$ versus the Borel parameter $M^2$,
   the $A$, $B$, $C$ and $D$ denote the   $D^*$, $D_1$, $B^*$ and $B_1$ mesons,  respectively. }
 \end{figure}

\begin{figure}
  \centering
  \includegraphics[totalheight=6cm,width=7cm]{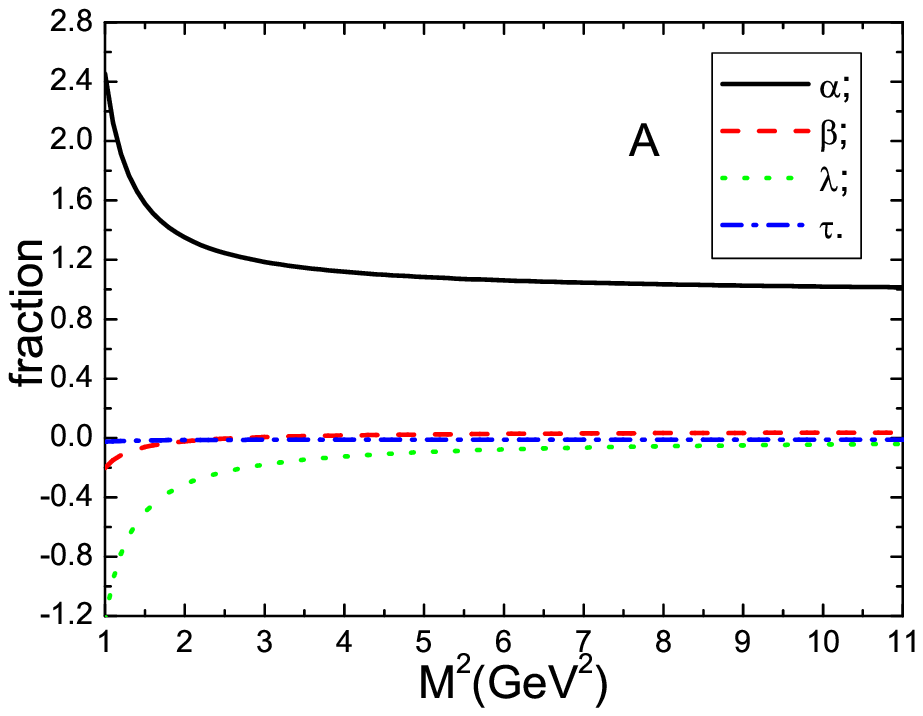}
  \includegraphics[totalheight=6cm,width=7cm]{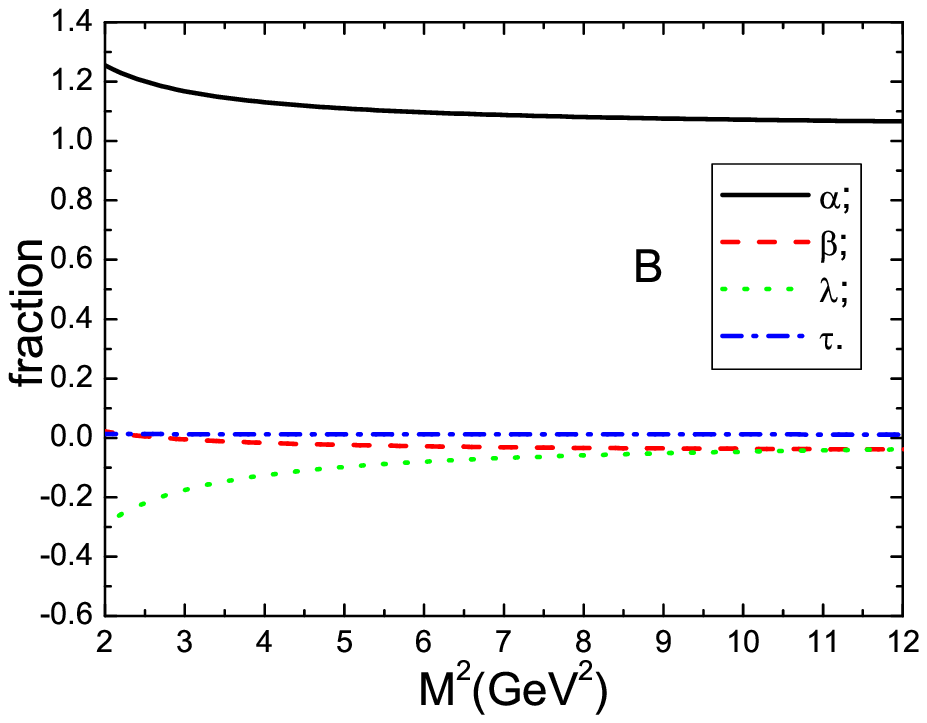}
   \includegraphics[totalheight=6cm,width=7cm]{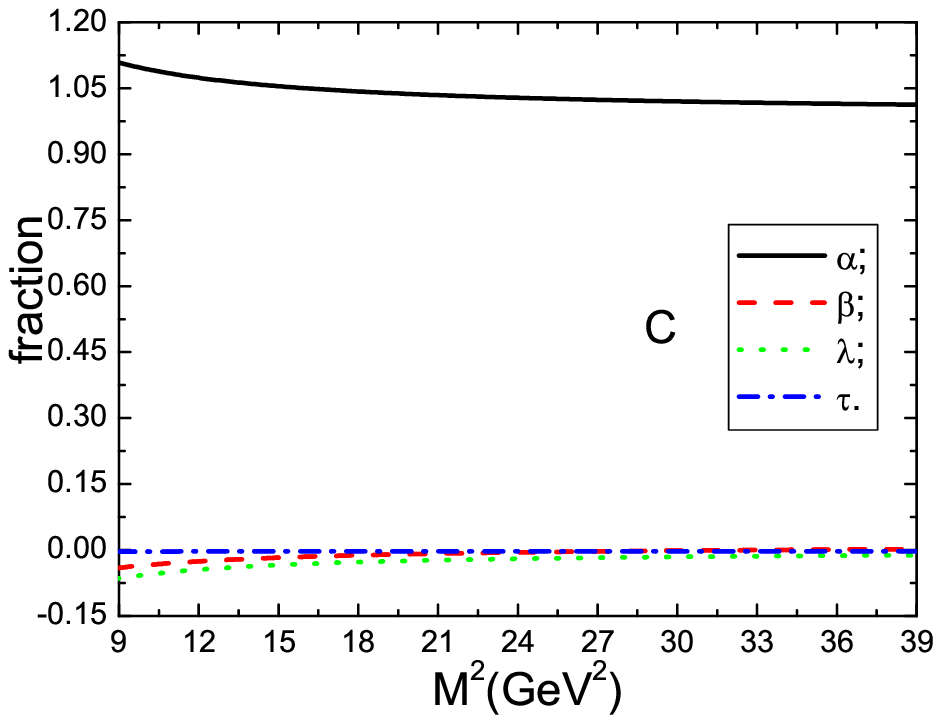}
    \includegraphics[totalheight=6cm,width=7cm]{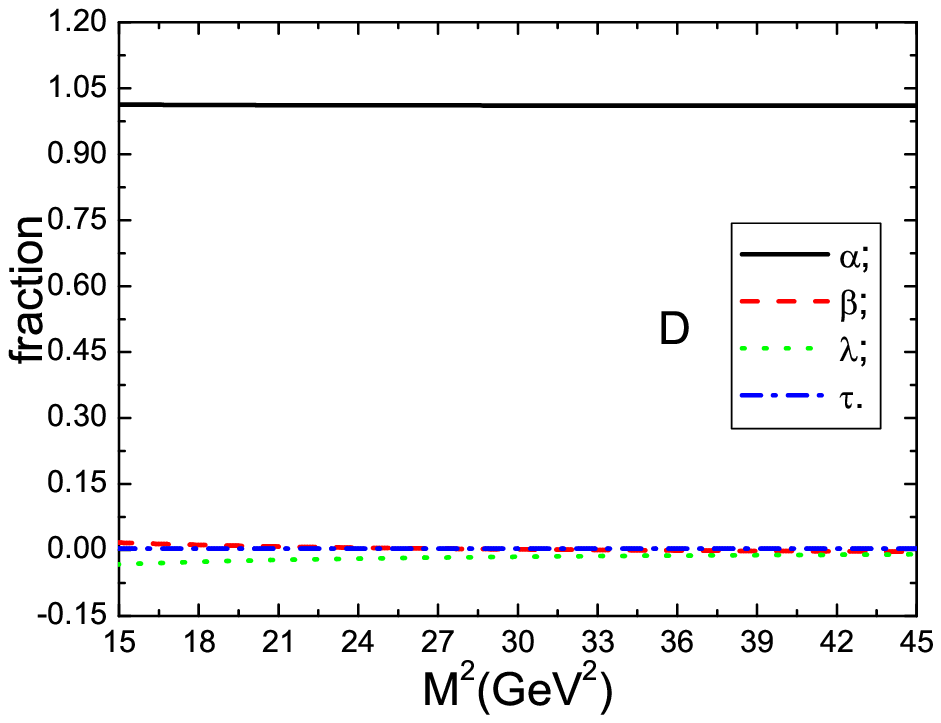}
  \caption{The   contributions from different terms versus the Borel parameter $M^2$ in the operator product expansion,
  where the $A$, $B$, $C$ and $D$ denote the   $D^*$, $D_1$, $B^*$ and $B_1$ mesons,  respectively; the $\alpha$, $\beta$, $\lambda$ and $\tau$
  denote the  $\langle\bar{q} q\rangle_N$, $\langle q^\dagger iD_0 q\rangle_N$, $\langle \bar{q} iD_0iD_0
q\rangle_N+ \langle\bar{q}g_s\sigma Gq\rangle_N $, and  $\langle\frac{\alpha_sGG}{\pi}\rangle_N$ terms, respectively. }
 \end{figure}

 \begin{figure}
  \centering
  \includegraphics[totalheight=6cm,width=7cm]{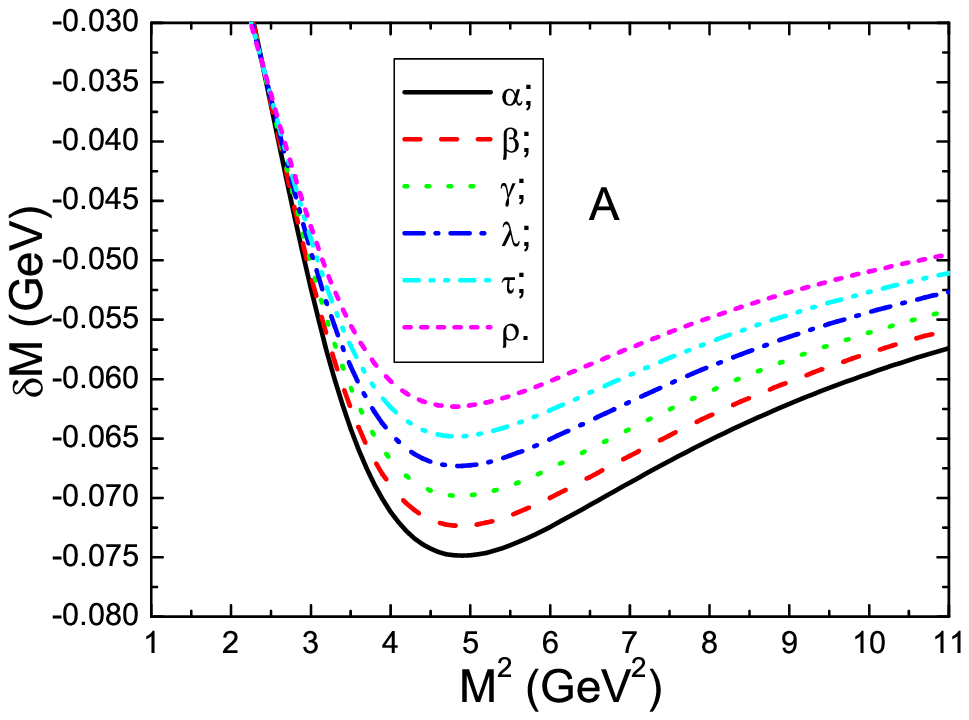}
  \includegraphics[totalheight=6cm,width=7cm]{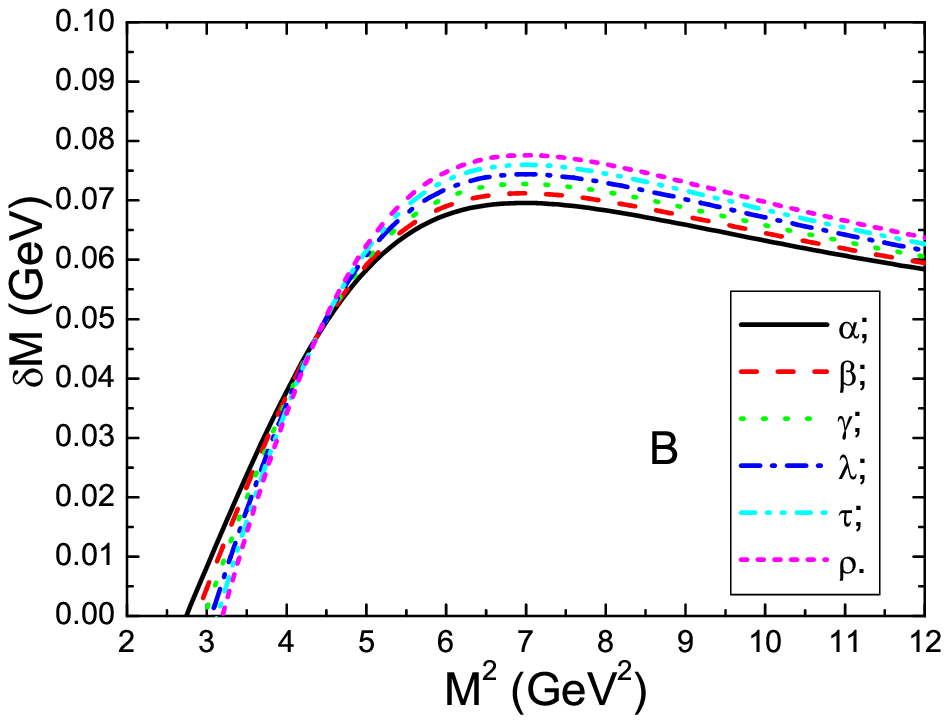}
   \includegraphics[totalheight=6cm,width=7cm]{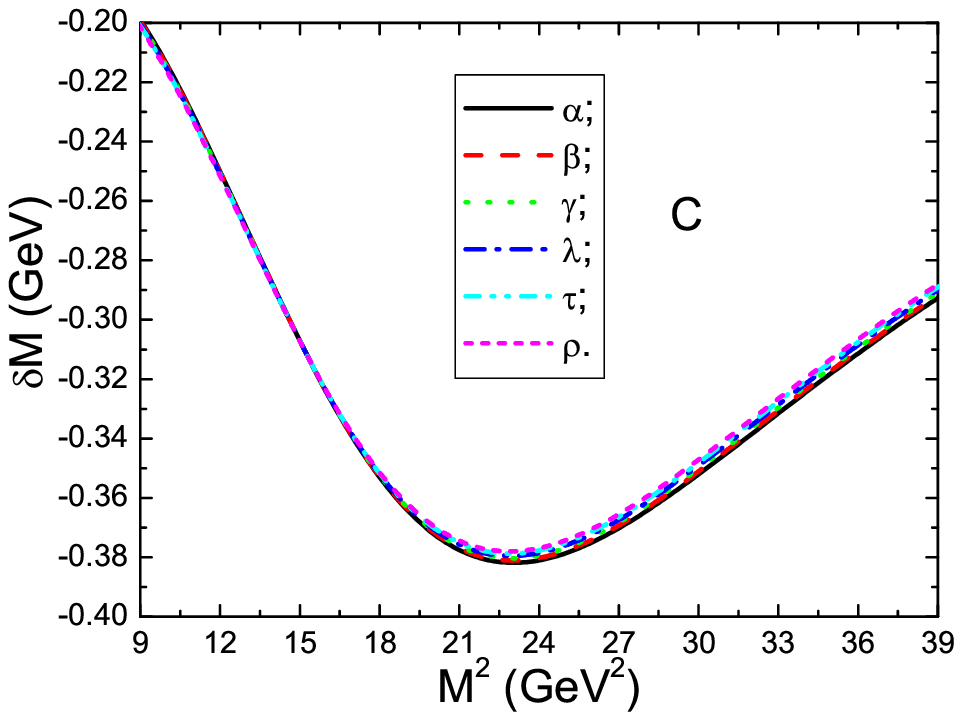}
    \includegraphics[totalheight=6cm,width=7cm]{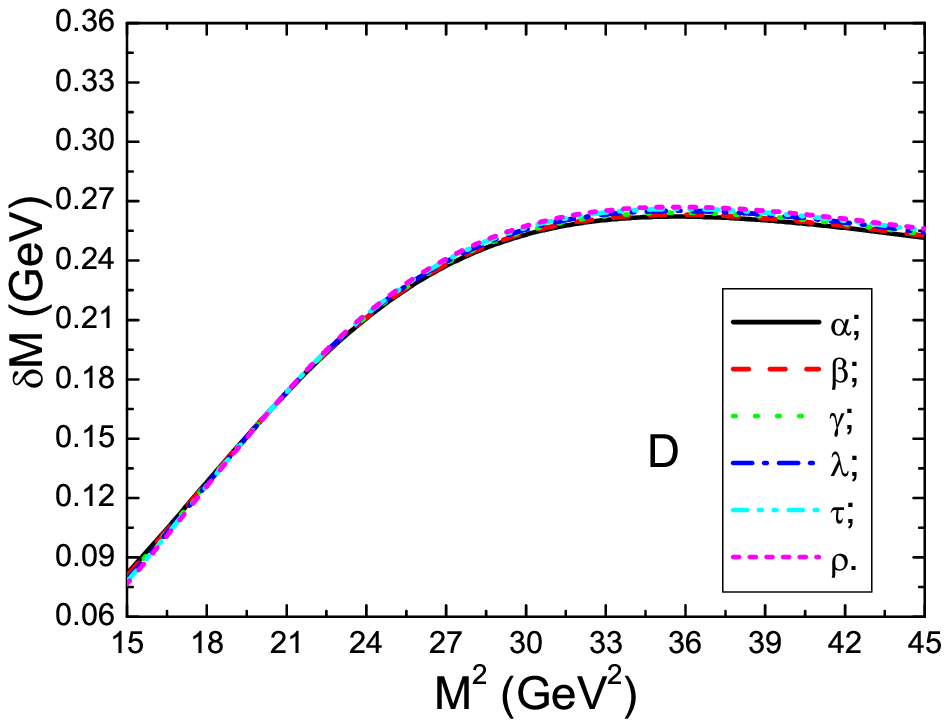}
  \caption{The   mass-shifts $\delta M$ versus the Borel parameter $M^2$,
   the $A$, $B$, $C$ and $D$ denote the   $D^*$, $D_1$, $B^*$ and $B_1$ mesons,  respectively; the $\alpha$, $\beta$, $\gamma$, $\lambda$, $\tau$
   and $\rho$ correspond the strong coupling constants $g^2=0$, $10$, $20$, $30$, $40$ and $50$, respectively. }
 \end{figure}

\section{Conclusion}
In this article, we calculate the  mass-shifts of the
vector and axialvector mesons $D^*$, $B^*$, $D_1$ and $B_1$ in the nuclear matter  using the QCD sum rules. We take the linear approximation at the low
density of the nuclear matter, and extract  the mass-shifts and scattering lengths explicitly,
$\delta M_{D^*}=-71\,\rm{MeV}$, $\delta M_{B^*}=-380\,\rm{MeV}$, $\delta M_{D_1}=72\,\rm{MeV}$,  $\delta M_{B
_1}=264\,\rm{MeV}$,  $a_{D^*}=-1.07\,\rm{fm}$, $a_{B^*}=-7.17\,\rm{fm}$,  $a_{D_1}=1.15\,\rm{fm}$ and $a_{B_1}=5.03\,\rm{fm}$.
 Our numerical results indicate that the  $D^*N$ and $B^*N$ interactions are attractive while the $D_1N$ and $B_1N$ interactions are repulsive;
 it is possible (difficult) to
form the $D^*N$ and $B^*N$ ($D_1N$ and $B_1N$) bound states.
    The $J/\psi$ production can obtain additional (no additional) suppression due to
 mass modification of the vector meson $D^*$ (axialvector meson $D_1$) in the  nuclear matter. The present predictions can be
 confronted with experimental data in the future.

\section*{Acknowledgements}
This  work is supported by National Natural Science Foundation, Grant Number 11075053,  and the Fundamental
Research Funds for the Central Universities.

\end{document}